\documentstyle[12pt]{article}
\begin{document}
\begin{titlepage}
{}~\vspace{1.5cm}

{}~\vskip1cm
\begin{center}
{\Large\bf Quark-Hadron Duality, Factorization and Strong Phases
in  $B^0_d\to \pi^+\pi^-$  Decay \\}
\vspace{1.0cm}
{\large I. Caprini and L. Micu\\[0.1cm]}
National Institute of Physics and Nuclear Engineering,\\
 POB MG 6, Bucharest, R-76900 Romania \\[0.3cm]
{\large C. Bourrely\\[0.1cm]}
Centre de Physique Th\'eorique 
\footnote{Laboratoire propre au CNRS-UPR 7061},CNRS-Luminy\\
Case 907, F-13288 Marseille Cedex 9 - France
\end{center}
\vspace{0.5cm}
\begin{abstract}
We consider the hadronic description of the  $B^0_d\to \pi^+\pi^-$ decay,
with the aim to investigate the strong phases generated by the final state
interactions. The derivation of the dispersion relations using the
Lehmann-Symanzik-Zimmermann formalism and the Goldberger-Treiman method to
include inelastic effects in the spectral function are presented.
We discuss the problem of quark-hadron duality and estimate in the hadronic
formalism the corrections to the factorized amplitude in the heavy
quark limit.
\end{abstract}
\vspace{1.0cm}
\end{titlepage}
\newpage
\section{Introduction}
The inclusion of the strong interaction effects in the theory of exclusive
nonleptonic $B$ decays  is a very difficult task.
The problem  has been investigated recently by many authors, in particular,
for charmless decays into light pseudoscalar mesons, since the strong phases of
these amplitudes are crucial for the determination of CP-violating phases in
present and future experiments \cite{Ball}. The first measurements of the
branching fractions of the $B$ decays into $\pi\pi$ and $\pi K$ final states
\cite{CLEO}-\cite{Belle}, considerably stimulated the theoretical and 
phenomenological  work devoted to these processes in various approaches. In 
the so-called "naive factorization approximation" \cite{BaSt}, the matrix 
elements of the operators entering the weak effective hamiltonian are 
expressed as products of meson decay constants and hadronic form factors, 
which are evaluated in a  phenomenological way. 
An obvious defficiency of this approximation is the renormalization scale 
dependence of the results, expressed as $\mu$-dependent Wilson coefficients  
multiplied by  $\mu$-independent hadronic form factors. 
Improvements to the factorization approximation  were discussed in 
several papers \cite{DuGr}-\cite{BeBu}. 
Recent calculations  of the $B\to \pi\pi$ decay amplitude were performed 
either in the
generalized QCD factorization approach \cite{BeBu}-\cite{BeBu2}, or by more
conventional perturbative QCD methods \cite{KeLi}-\cite{LuUk}. 

The nonleptonic $B$ decays were also investigated recently in a hadronic
approach, in which a part of the strong dynamics accompanying the weak decay is
described using the unitarity of the $S$-matrix, dispersion relations and 
Regge phenomenology \cite{Dono}-\cite{Zenc}. In the present paper, 
we apply this approach to the particular case of $B^0_d\to \pi^+\pi^-$ 
decay. One aim of our study is to compare the predictions of the hadronic and 
the partonic treatments and to test the validity of quark-hadron duality. 

In the next section, we discuss the derivation of dispersion relations  
with respect to the momentum squared of the external particles, by applying the
Lehmann-Symanzik-Zimmermann (LSZ) reduction  formalism \cite{LSZ} to the
$S$-matrix element  of the decay process. In Sect. \,\ref{sec3} we explain the 
Goldberger-Treiman  procedure to include inelastic contributions in the
spectral function and apply it to the amplitudes of the decay $B^0_d\to
\pi^+\pi^-$. In Sect.\,\ref{sec4} we consider the problem of quark-hadron duality, 
and estimate in the  hadronic formalism the corrections
to the factorized amplitude produced by the final state interactions in the
heavy quark limit. Section  \,\ref{sec5}  contains our conclusions.
 
\section{Dispersion relations for the decay amplitude}\label{sec2}
We consider the decay amplitude
\begin{equation}\label{defa1}
A (B^0_d\to\pi^+\pi^-)=  \langle \pi^+(k_1)\, \pi^-(k_2), {\rm out}|{\cal
H}_w(0)|B^0_d(p), {\rm in}\rangle\,, 
\end{equation}
where the ``in'' and ``out'' states  are defined with respect to the strong 
interactions and ${\cal H}_w$ is the weak effective hamiltonian density
\begin{eqnarray}\label{Hw} 
&&{\cal H}_w={G_F\over \sqrt{2}}\sum\limits_{j=u,c} V_{jd} V^*_{jb}  \\
&&\times\left[
C_1(\mu) O_1^j(\mu)+C_2(\mu)O_2^j (\mu) +\sum\limits_{i=3,\ldots ,8} C_i(\mu)
O_i(\mu)\right]\,.\nonumber
\end{eqnarray}   
In this relation, $O_i$ are local $\Delta B=1, \Delta S=0$ operators, and
$C_i$  the corresponding Wilson coefficients, which take into account
perturbatively the strong dynamics at distances shorter than $1/\mu$. 
Using the expression  (\ref{Hw}) of the weak hamiltonian, the decay amplitude
(\ref{defa1}) can be splitted in two terms  
\begin{equation}\label{Auc}
A(B^0_d\to \pi^+\pi^-) =V_{ud} V^*_{ub} A_u + V_{cd} V^*_{cb} A_c \,,
\end{equation} 
with the CP violating phase $\gamma=\mbox{Arg}(V^*_{ub})$appearing in the 
first term.

The physical amplitude (\ref{defa1}) is calculated for $p=k_1+k_2$ at on-shell
values of the momenta, $p^2=m_B^2$, $k_1^2=m_\pi^2$, $k_2^2=m_\pi^2$.
The extrapolation to off-shell external momenta can be achieved by the LSZ
reduction formalism \cite{LSZ}. In Refs. \cite{CaMi} we applied this technique
to the expression (\ref{defa1}) of the amplitude. 
As we shall see below, it is more convenient to start from the $S$-matrix 
element
\begin{equation}\label{defS} 
S_{B^0_d\to\pi^+\pi^-}=\langle \pi^+(k_1)\,
\pi^-(k_2), {\rm out}| B^0_d(p), {\rm in}\rangle\,, 
\end{equation} 
where the transition from the ``in'' to the ``out'' states includes
both the strong and weak interactions. 
The expression (\ref{defa1}) of the decay amplitude is
obtained by expanding the $S$ matrix to first order in the weak
hamiltonian. However, we can apply the LSZ reduction to the $B$ 
meson in Eq. (\ref{defS}), which leads to the alternative expression 
\begin{equation}\label{defa2} 
A (B^0_d\to\pi^+\pi^-)= {1\over \sqrt{2
p_0}}\langle \pi^+(k_1)\, \pi^-(k_2), {\rm out}| \eta_{B^0}(0)|0\rangle\,,
\end{equation}
where $\eta_{B^0}(x)={\cal K}_x \phi_{B^0}(x)$ is the source of the meson 
$B^0_d$ and $\phi_{B^0}$ its
interpolating field  (${\cal K}_x $ is the Klein-Gordon operator).
We recall that in a Lagrangian theory the source, which includes both the
strong and weak interactions, can be written formally as
\begin{equation}\label{etaB} 
\eta_{B^0}(x)={\delta {\cal L}_{int}\over \delta 
\phi_{B^0}}-\partial_\mu{\delta 
{\cal L}_{int}\over \delta \partial_\mu\phi_{B^0}}.
\end{equation}
The matrix element in Eq. (\ref{defa2}) can be defined for arbitrary
$s=p^2=(k_1+k_2)^2$, the physical amplitude corresponding to $s=m_B^2$.
We notice that Eq. (\ref{defa2}) is similar to the definition of the pion
electromagnetic form factor, where $\eta_B$ is replaced by the electromagnetic
current $J_\mu$. We can apply therefore the standard methods used in deriving
the dispersion relations for the pion form factor \cite{GoWa}-\cite{Binc}. 
More precisely, by the LSZ reduction of one final meson (say, $\pi^+$) 
in Eq. (\ref{defa2}), we obtain
\begin{eqnarray}\label{lszas} 
&&A (B^0_d\to\pi^+\pi^-) = {i\over \sqrt{4
k_{10} p_0}}\int {\rm d}x e^{ik_1 x} \theta (x_0) \\ 
&&\times\langle \pi^-(k_2)|[\eta_{\pi^+}(x),
\eta_{B^0} (0)]|0\rangle\,
-{i\over \sqrt{4 k_{10} p_0} }\int {\rm d}x
e^{ik_1 x} \delta(x_0) \nonumber \\
&&\times\langle \pi^-(k_2)|i k_{10}[\phi_{\pi^+}(x),
\eta_{B^0}(0)]\,- \, [\partial_0\phi_{\pi^+}(x),
\eta_{B^0}(0)]|0\rangle\,,\nonumber 
\end{eqnarray}
where $\eta_{\pi^+}(x)$ is the source of the reduced  pion.

The second integral in Eq. (\ref{lszas}) contains equal-time commutators
produced by the action of the Klein-Gordon operator
${\cal K}_x$ upon the function $\theta(x_0)$. As shown in \cite{Bart}, the
most general form of this term, called "degenerate", is a constant or a
polynomial of the Lorentz invariant variables.
To calculate the degenerate term, one needs the expression of the source
$\eta_B$ in terms of  the interpolating field $\phi_{\pi^+}$
or its time  derivative, which in a hadronic Lagrangian theory might
be obtained from the formal expression (\ref{etaB}). 
The commutators can be then evaluated in principle by applying the canonical
commutation rules, satisfied,  up to a normalization constant, by the
interpolation fields \cite{Bart}. However, in the standard model, the hadronic
fields are defined in terms of the underlying quark and gluon degrees of
freedom, and the definition of the off-shell fields might introduce 
ambiguities in the evaluation of the degenerate term (the result depends also
on which pion, $\pi^+$ or $\pi^-$, is reduced, since their quark content is
different).

We now turn to the first term of Eq. (\ref{lszas}), which is usually called 
"dispersive term", and has a more complicated structure as a  function of the 
squared external momenta. The integral defines a function holomorphic at the 
values of the momenta for which it is convergent.   
First, due to the presence of the $\theta(x_0)$ function, the
integral upon $x_0$ in Eq.  (\ref{lszas}) is convergent for ${\rm Im}
k_{10}>0$, {\em i.e.} in the upper half of the $k_{10}$ complex  plane.  A
detailed analysis must exploit also the causality properties of the commutator
\cite{Bart}, which restricts the integral upon the spatial variables to $|{\bf
x}|< |x_0|$. The difficult part of the conventional proofs of the
dispersion relations is to show  that the integrals upon $x_0$ and ${\bf x}$ 
are convergent for complex values of the external  momenta.  
As discussed in \cite{Bart}, it is sometimes useful to go 
to a particular Lorentz frame and consider a particular variable, 
for instance $k_{10}$, instead of trying to think in terms of Lorentz 
invariants. Also, it is useful to treat simultaneously the matrix elements
$\langle\pi\pi, {\rm out}|\eta_B|0\rangle, \langle\pi\pi, {\rm
in}|\eta_B|0\rangle$ and $\langle\pi |\eta_B|\pi\rangle$, which are
represented by the same analytic function in various parts of the
complex plane of the dispersive variable.  

In the present case, it is convenient to choose the system with the
unreduced pion $\pi^-$ at rest (${\bf k_2}=0$), when
$k_{10}=(s-k_1^2-m_\pi^2)/ 2 m_\pi$. In what follows we shall either work 
with $s$ variable keeping  $k_1^2=m_\pi^2$ fixed, or with $k_1^2$
variable  at fixed $s=m_B^2$.   By expressing ${\bf k}_1^2$  in
terms of $k_{10}$ and the fixed Lorenz invariant momentum, the first
term of Eq.  (\ref{lszas}) depends only on the variable $k_{10}$, and
is analytic in the complex $k_{10}$ plane, except a possible 
discontinuity along the real axis, given by \cite{Bart}
\begin{eqnarray}\label{spectra} && \sigma
(k_{10}) = {1\over 2 \sqrt{4 k_{01}p_0}}
\nonumber\\&&\times\left[\sum_{n}\delta  (k_1+k_2-p_n) \langle \pi^-(k_2)|
\eta_{\pi^+}(0)|n\rangle\langle n| \eta_{B^0}(0)|0\rangle\right.\nonumber \\
&&- \left.\sum_{n}\delta  (k_1+p_n)  \langle \pi^-(k_2)|
\eta_{B^0}(0)|n\rangle\langle n| \eta_{\pi^+}(0)|0\rangle\right]\,.
\end{eqnarray}   
This expression is obtained formally from (\ref{lszas}) by
replacing $i\theta(x_0)$ by $1/2$, inserting a complete set of intermediate
states in the commutator and using translational invariance \cite{Bart}. In
order to evaluate the spectral function, we recall that the sources contain
both the strong and the weak  interactions, the last ones being treated to
first order. As we shall see below, the spectral function takes different
forms, depending on the external momentum adopted as dispersive variable.

Let us assume first that $k_1^2$ is fixed at the physical
value $k_1^2=m_\pi^2$, and treat the amplitude as a function of the variable
$s=(k_1+k_2)^2$.  In the reference system chosen above
$k_{10}=(s-2m_\pi^2)/ 2 m_\pi$ and ${\bf k}_1^2=k_{10}^2-m_\pi^2$. It is easy
to see that in this case the  second sum in the expression (\ref{spectra})
brings no contribution. Indeed, since $k_1^2=m_\pi^2$, the only state which
contributes is the one-pion state $|n\rangle=|\pi\rangle$, and $\langle
\pi|\eta_\pi|0\rangle=0$ \cite{Bart}.

As concerns the first sum of Eq. (\ref{spectra}), the intermediate states $n$ 
which contribute are of two kinds: the first ones are generated by the weak
part of the source  $\eta_B$ in the second matrix element, and undergo a strong
transition to the final state $\pi^+\pi^-$, mediated   by the strong part of
$\eta_\pi$. According to Eq. (\ref{defa2}), the second matrix element is equal
to the weak decay amplitude of an off-shell meson $B$, with momentum squared
equal to $s$, while the first matrix element is a strong amplitude, evaluated
at the c.m. momentum squared equal to $s$. Therefore, the contribution of
these states in the unitarity sum represents the so-called  "final state
interactions" (FSI). The lowest intermediate state consists of
two-pions, which produces the lowest branch-point at $k_{10}=m_\pi$, or,
equivalently, $s=4m_\pi^2$. 

The intermediate states $n$ of the second type are produced by the strong part 
of the source $\eta_B$ in the second matrix element, which describes the 
strong  decay of an off-shell $B$ meson. 
The first matrix element, where  contributes the weak part of $\eta_\pi$,
describes the weak transition amplitudes from the state $n$ to the final
$\pi^+\pi^-$ state. These terms  are usually interpreted as
"initial state interactions" (ISI). The lowest physical state which can
contribute is the pair $B^* \pi$, producing the lowest threshold
$s=(m_{B^*}+m_\pi)^2$.

The whole amplitude can be recovered from the discontinuity by means of a
dispersion integral. To write it down, we need the asymptotic behaviour of the
discontinuity, which is difficult to estimate, since it involves off-shell
quantities.  Assuming, for simplicity, that one subtraction is necessary, and
combining the possible degenerate terms, discussed above, with the subtraction
constant of the dispersive part, we express the physical amplitude as 
\begin{eqnarray}\label{inta1}  
&&A(B^0\to\pi^+\pi^-)= A (s_0)\nonumber \\&& +
{m_B^2- s_0\over \pi }\int\limits_{4 m_\pi^2}^\infty {\rm d}s{\sigma_{FSI} (s)
 \over (s-m_B^2-i\epsilon) (s-s_0)}\,\nonumber \\  &&+ {m_B^2-s_0\over \pi
}\int \limits_{(m_B^*+m_\pi)^2}^\infty   {\rm d}s{\sigma_{ISI}(s)\over
(s-m_B^2) (s-s_0)} \,, 
\end{eqnarray}  
where 
\begin{eqnarray}\label{sigfi}
\sigma_{FSI}\approx \sum\limits_n \delta(k_1+k_2-p_n) M^* (n\to \pi\pi)
A(B\to n)\,\nonumber\\ \sigma_{ISI}\approx \sum\limits_n
\delta(k_1+k_2-p_n) A^*(n\to \pi\pi) M (B\to n) 
\end{eqnarray}  
are the spectral functions associated to the final (initial) state
 interactions,
respectively. In these relations, $A(M)$ denote the amplitudes of the weak
(strong) transitions, respectively, evaluated for an off-shell $B$ momentum
squared equal to $s$. Since $B$ is stable with respect to the strong
interactions, $m_B < m_{B^*}+m_\pi$, the initial state interactions (the last
integral in Eq. (\ref{inta1})) do not contribute to the on-shell imaginary
part. We mention that a dispersion relation  similar to Eq. (\ref{inta1}) was
derived recently in Ref. \cite{Suzu} for $K\to \pi\pi$ decay, starting from
the definition (\ref{defa1}) of the decay amplitude, treating  the weak
hamiltonian ${\cal H}_w$ as the source of a spurion, and using the
Mandelstam representation. 

As mentioned above, the expression (\ref{lszas}) can be analytically continued
also in the variable $k_1^2$, at fixed $s$, equal to  the physical value 
$s=m_B^2$. In this case, in the reference system chosen above, 
$k_{10}=(m_B^2-k_1^2-m_\pi^2)/ 2 m_\pi$ and
${\bf k}_1^2= [k_1^2-(m_B+m_\pi)^2][k_1^2-(m_B-m_\pi)^2]/(2m_\pi)^2$.  
The spectral function is given formally by the same expression (\ref{spectra}),
but now the contributions are  different. First, we notice that the second sum
in Eq. (\ref{spectra}), which previously vanished on account of
$k_1^2=m_\pi^2$, now includes terms which are produced by the strong and the
weak parts of the source $\eta_\pi$, for a variable $k_1^2$. Just like in the
discussion above, there are intermediate states which are generated by the
strong decay of an off-shell pion, and undergo then a weak transition
mediated by the weak part of $\eta_B$, and also
intermediate states which  are produced by the weak part of $\eta_\pi$ and
generate afterwards a pion and a $B$ meson through a strong interaction. In
the first case the lowest branch point is at $k_1^2=9 m_\pi^2$,
corresponding to the intermediate state with three pions, and in the second
case at $k_1^2=(m_B+m_\pi)^2$, corresponding to the  intermediate state 
$\pi B$. From the connection between  $k_{10}$ and $k_1^2$ in the particular
system mentioned above, one can see that the branch cut corresponds to
negative values of $k_{10}$. This means that these contributions originate
actually  from the matrix element $\langle \pi |\eta_B|\pi\rangle$, related to
the $B\to\pi\pi$ decay amplitude by crossing symmetry.

As concerns the first sum in the spectral function (\ref{spectra}), the
intermediate states which bring a nonvanishing contribution have
$p_n^2=(k_1+k_2)^2$ fixed at the value $m_B^2$. The strong part of $\eta_B$
gives therefore no contribution, since the lowest state possible (the pair
$B^* \pi$) can not be produced at this energy. On the other hand, the weak
part of the source $\eta_B$ brings a nonvanishing contribution, producing
intermediate particles which undergo then a strong interaction. 
This contribution represents therefore the final state interactions. 
The delta function implies the condition  $k_1^2=(p_n-k_2)^2$, where
$p_n^2=m_B^2$ and  $k_2^2=m_\pi^2$, which gives for $k_1^2$ the allowed range
$k_1^2\le (m_B-m_\pi)^2$. We notice that, unlike the other branch points
discussed above, which are determined by the lowest intermediate states
in the unitarity sum, the range of the variable $k_1^2$ has a more  
kinematical nature.

We express now the whole amplitude in terms of its discontinuity by a
dispersion integral. Assuming that one subtraction is necessary and
including the degenerate terms in the subtraction constant, we obtain   
\begin{eqnarray}\label{inta2} 
&&A(B^0\to \pi^+\pi^-)= A(\kappa_0^2) \\
&&+{(m_\pi^2-\kappa_0^2)\over \pi} 
\int\limits_{-\infty}^{(m_B-m_\pi)^2}{\rm d} k_1'^2 {\sigma_{FSI}
(k_1'^2)\over (k_1'^2 -m_\pi^2-i\epsilon)(k_1'^2-\kappa_0^2)}\nonumber \\  
&&+{(m_\pi^2-\kappa_0^2)\over \pi} 
\int\limits^\infty_{9m_\pi^2}{\rm d} k_1'^2 {\sigma_{1} (k_1'^2)\over (k_1'^2
-m_\pi^2)(k_1'^2-\kappa_0^2) }\nonumber\\
&&+{(m_\pi^2-\kappa_0^2)\over \pi}  
\int\limits^{\infty}_{(m_B+m_\pi)^2}{\rm d} k_1'^2 {\sigma_{2} (k_1'^2)\over
(k_1'^2 -m_\pi^2)(k_1'^2-\kappa_0^2) }\,,\nonumber 
\end{eqnarray} 
where $\sigma_{FSI}$ is given formally by the same expression
(\ref{sigfi}) given above, evaluated now at fixed $s=m_B^2$ and variable
$k_1^2$, and 
\begin{eqnarray}\label{sigma12} &&
\sigma_{1}
\approx \sum\limits_{n=3\pi, \ldots} \delta(k_1+p_n) A^*(n\to \pi B) M(\pi \to
n) \nonumber\\&&
\sigma_{2}\approx \sum\limits_{n=B\pi,\ldots}
\delta(k_1+p_n) M^*(n\to \pi B) A(\pi \to n),. 
\end{eqnarray}
We denoted generically, as before, by $A (M)$ the weak (strong) amplitudes,
respectively.

It is easy to see that both dispersion relations (\ref{inta1}) and 
(\ref{inta2})  lead to the same discontinuity on shell, equal to the spectral
function 
$\sigma_{FSI}$ evaluated for physical masses. These relations might be
useful in principle if the decay amplitude can be calculated  (by chiral
theory, lattice, etc) at some particular points ($s=s_0$ or
$k_1^2=\kappa_0^2$), with a better accuracy than at the physical points,
$s=m_B^2$ and $k_1^2=m_\pi^2$, respectively. The complete evaluation of 
the dispersion relations is very difficult, since they involve off-shell
quantities in the spectral functions.  However, as we will show below, the 
dispersion relation (\ref{inta2}) can be written in a different form,
more convenient for the study of the final state interactions. We first remark
that the spectral function $\sigma_{FSI}$ defined in Eq. (\ref{sigfi})
is actually independent of the dispersive variable
$k_1^2$, for fixed $(k_1+k_2)^2=m_B^2$. This can be easily seen by performing 
the phase space integral  in the first sum of Eq. (\ref{spectra}) in the
c.m. system, when ${\bf p}_n={\bf k_1}+{\bf k}_2=0$ and the energy squared is
equal to $m_B^2$. As the weak decay amplitude (the matrix element 
$ \langle n|\eta_B|0\rangle/\sqrt{2 p_{0}}$  in Eq.  (\ref{spectra})) and the
invariant strong amplitude (the matrix element 
$\langle \pi^-(k_2)| \eta_{\pi^+}(0)|n\rangle/\sqrt{2 k_{01}}$)  
depend both only on the Mandelstam variables and the  physical masses of the 
particles involved, the result of the phase space integration at fixed $s$ 
is independent of  $k_1^2$ and contains only on-shell
quantities\footnote{The unusual property of the spectral function
to be independent of the dispersive variable $k_1^2$, was
noticed a long time ago \cite{Bart}.}.

We notice that the most general form of a function having a branch cut for
$-\mu^2\le k_1^2\le (m_B-m_\pi)^2$, with a constant discontinuity
$\sigma_{FSI}$, is  \begin{equation}\label{disp} 
A_{FSI}(k_1^2) =  {\cal P}(k_1^2)+ 
{\sigma_{FSI}\over\pi}  \ln\left[{k_1^2 -(m_B-m_\pi)^2 \over
\mu^2+k_1^2}\right] , \end{equation}   
where ${\cal P}(k_1^2)$ is a polynomial (more generally, an entire function),
independent on  $\sigma_{FSI}$. In order to construct the
full decay amplitude  we must add to the function $A_{FSI}$ the contribution
of the degenerate terms and the last two dispersion integrals in Eq.
(\ref{inta2}), with possible subtractions. All these terms, as well as the
polynomial ${\cal P}(k_1^2)$, are independent of the discontinuity
$\sigma_{FSI}$. Therefore, by combining them into a single constant, and
choosing  the scale $\mu^2\approx m_B^2$,  we write the physical amplitude as 
\begin{eqnarray}\label{direla} A(B^0_d\to
\pi^+\pi^-) = A_0 + {\sigma_{FSI}\over\pi} \ln\left[{m_\pi^2\over
(m_B-m_\pi)^2}-1\right] 
\end{eqnarray}
where $A_0$ is the genuine contribution which remains  when the long
distance final  state interactions are switched-off, {\em i.e.}
$\sigma_{FSI}=0$.  It is worth mentioning that this separation of the final
state interactions  from the other parts of the dynamics was possible only due
to the fact that  $\sigma_{FSI}$ does not depend on the dispersive variable,
allowing us to  construct $A_{FSI}$ according to (\ref{disp}). 
The subtraction constants in the usual subtracted  dispersion
relations, like Eqs. (\ref{inta1}) or (\ref{inta2}), do not have a similar
interpretation: they represent the values of the amplitude at some particular
points, and depend implicitly on all the spectral function in the dispersion
relation.

We shall end this section with some comments. First, we recall that rigorous 
analytic properties in the external momenta are proved in
axiomatic field theory  only for a small region close to the physical
masses \cite{KaWi}. Therefore, the  dispersion relations presented above 
can be  accepted only as an heuristic conjecture, whose validity remains 
to be tested.  
We mention also that  the  off-shell analytic continuation in external momenta 
is in general plagued by ambiguities.  
They may appear, in the present formalism, in the evaluation of 
the degenerate terms and of the off-shell amplitudes entering the
spectral functions. Moreover, we notice that even the 
analytic properties of the off-shell amplitude may depend on the
specific expression of the on-shell amplitude, used as starting point of the
extrapolation. For instance,  by applying the LSZ
procedure to the expression (\ref{defa1}) of the amplitude, we obtained in Ref.
\cite{CaMi}  only a part of the dispersive branch-cuts  written in Eqs.
(\ref{inta1}) and  (\ref{inta2}).  The contribution of the
missing dispersion integrals (namely, the FSI contribution in Eq. (\ref{inta1})
and the ISI contribution in Eq. (\ref{inta2})) is hidden in the
corresponding degenerate terms, which have a different form \cite{CaMi}. Of
course, one expects that the amplitude on-shell is recoverd in an univoque
way, but the compensation of the ambiguities of various terms is difficult
to see in approximate calculations.   

As concerns the phenomenological applications,    
the final state interactions in $B$ hadronic decays were
investigated the last  years  by means of dispersion relations with respect to
the momentum squared ($s$) of the $B$ meson \cite{BlHa}-\cite{Falk} (recently,
this method was applied also to $K\to \pi\pi$ decay \cite{Suzu}-\cite{BuCo}).
These dispersion relations  look more familiar, due to their formal
resemblance with the case of the pion form factor. However, as seen from Eq.
(\ref{inta1}), the similarity is not complete, due to the presence of the
initial strong interactions in the weak decay and the appearance of off-shell
quantities. The dispersion relations  with respect to the momentum squared 
$k_1^2$, which seem less intuitive, were applied to $B$ decays in Refs.
\cite{CaMi} (we mention also one earlier application of this technique
for the calculation of the nucleon form factor \cite{Binc}). In Section 3
we shall use this type of dispersion relations, written in the convenient form
(\ref{direla}), for discussing the  effects of the final state  interactions
in the $B^0\to\pi^+\pi^-$ decay. Before making this analysis, we shall first
investigate the  spectral function $\sigma_{FSI}$ appearing in this relation.

\section{The Goldberger-Treiman procedure} \label{sec3}
Following Ref. \cite{BeBu}, we shall use the parametrization
\begin{eqnarray}\label{abbns}&& A(B^0_d\to
\pi^+\pi^-) \,= i{G_F\over \sqrt{2}}m_B^2 f_+(m_\pi^2)f_\pi|V_{ud}V^*_{ub}|
{\rm e}^{i\gamma} \nonumber \\ &&\times \left[T_u(B\to  \pi^+\pi^-) +{{\rm
e}^{-i\gamma}\over R_b}\,T_c (B\to \pi^+\pi^-)\right]\,, 
\end{eqnarray}
obtained by extracting from the amplitudes $A_u$ and $A_c$ of 
Eq. (\ref{Auc}) the "naive" factorized amplitude, 
expressed in terms of the pion decay constant $f_\pi$ and the
$B\to \pi$ transition form factor $f_+(m_\pi^2)$ ( $R_b=|V_{ub}/(\lambda
V_{cb})|(1-\lambda^2/2)\approx 0.377$). The spectral function $\sigma_{FSI}$
defined in Eq. (\ref{sigfi}) can be written in a similar way as
\begin{equation}\label{sigmaT} 
\sigma_{FSI} =  i{G_F\over \sqrt{2}}m_B^2
f_+(m_\pi^2)f_\pi |V_{ud}V^*_{ub}| {\rm e}^{i\gamma} \left[\sigma_{FSI}^u +
{{\rm e}^{-i \gamma}\over R_b}\sigma_{FSI}^c \right]\,, 
\end{equation} 
where, recalling that $\sigma_{FSI}$ is given by the first term in the 
unitarity sum (\ref{spectra}), we have
\begin{eqnarray}\label{sigmaj}  
\sigma_{FSI}^j \sim  \sum_{n}\delta
(k_1+k_2-p_n)&&\langle \pi^-| \eta_{\pi^+}|n\rangle\langle n|
\eta_B^j|0\rangle,\nonumber\\ &&j=u,c\,. 
\end{eqnarray}  
We assumed here that the source $\eta_B$ admits a decomposition in two terms, 
analogous to that of the weak Hamiltonian (\ref{Hw}). 
This shows also that one can derive separate dispersion relations for each 
of the amplitudes $T_u$ and $T_c$.
In particular, according to Eq. (\ref{direla}), we shall write these 
relations in the form 
\begin{equation}\label{direlT} T_j = T_{j,0}+
{\sigma_{FSI}^j\over\pi}  \ln\left[{m_\pi^2\over (m_B-m_\pi)^2}-1\right]
\,,\quad j=u,c,
\end{equation} 
where we denoted by  $T_{j,0}$ the analog of the term $A_0$ in the relation 
(\ref{direla}), divided by the constant factorized in (\ref{abbns}). 

For further applications, it is important to notice that the spectral functions
$\sigma^u_{FSI}$ and $\sigma_{FSI}^c$ are real quantities:    
\begin{equation}\label{reality} 
\sigma_{FSI}^j =(\sigma^{j}_{FSI})^*\,,\quad
j=u,c \,. 
\end{equation} 
The proof of these equalities is based on the properties of the matrix 
elements in Eq. (\ref{sigmaj}) under the $PT$ transformation \cite{Bart}. 
More precisely, in the present case, we have
\cite{CaMi} 
\begin{eqnarray}\label{PT} 
\langle  \pi^-(k_2)| 
\eta_{\pi^+}(0)|n, {\rm in}\rangle &=& \langle  \pi^-(k_2)| 
\eta_{\pi^+}(0)|n, {\rm out}\rangle^*\nonumber \\
\langle n, {\rm in}| \eta_B^j (0) |0 \rangle &=&- \langle n, {\rm out}|
\eta_B^j(0) |0\rangle^*\,,
\end{eqnarray}  
where the minus sign in the second relation is due to the specific 
spin-parity properties of the relevant part of the weak hamiltonian. 
Using the equalities (\ref{PT}) in Eq. (\ref{sigmaj}), and
taking into account the equivalence of the complete sets of "in" and "out"
intermediate states in the unitarity sum, it is easy to prove
the relations (\ref{reality}) (the minus sign in Eq. (\ref{PT}) is compensated 
by leaving aside an imaginary constant in the definition (\ref{sigmaT})). 

In approximate calculations, the set of intermediate states in the unitarity
sum (\ref{sigmaj}) is truncated, which might lead to violations of the
reality conditions (\ref{reality}), and to  the appearance of artificial strong
phases in the spectral functions. This fact is important in the present case,
since the unitarity sum  is evaluated at large c.m. energy squared, 
$s=m_B^2$,  where many inelastic channels are open. 
However, with a ``good'' choice of the truncated set one can avoid the
appearance of unphysical phases.
The idea of Goldberger and Treiman \cite{GoTr} was to take the
intermediate states in the symmetric combination 
$1/2 |n, {\rm in}\rangle\langle n, {\rm in}| +1/2 |n, {\rm
out}\rangle \langle n, {\rm out}|$,

 which represents also a complete set.
The remarkable point is that, even when it is truncated, this set generates
spectral functions  which satisfy the reality condition (\ref{reality}), at
each step of approximation. The symmetric summation simulates therefore,
in a certain measure, the effects of inelastic states, without incorporating 
them explicitly in the unitarity sum\footnote{An alternative approach to
include the effects of the inelastic states in $B$ hadronic decays, based on 
statistical arguments and Regge phenomenology, was proposed in 
Refs.\cite{SuWo}, \cite{Zenc}}.

It is worth mentioning that a complete set written in a symmetric form is 
quite natural in the LSZ method: indeed, when deriving
the discontinuity of the amplitude, the initial $\theta(x_0)$
function in Eq. (\ref{lszas}), whose origin is the reduction of an
"out" pion, is actually replaced by $\theta(x_0)/2+\theta(-x_0)/2$
\cite{Bart}. This means that  the final two-pion state appears in the 
discontinuity  in the symmetric combination  
$1/2 |\pi^+ \pi^-, {\rm out}\rangle + 1/2 |\pi^+ \pi^-, {\rm in}\rangle$
\footnote{A similar symmetric combination of "in" and
"out" states was obtained in a related context in Ref. \cite{MaTe}.}. 
It is therefore reasonable to take the same symmetric combination 
also for the intermediate states $n$.

The Goldberger-Treiman procedure allows us to write the spectral functions
$\sigma_{FSI}^j$ defined in Eq. (\ref{sigmaj}) as
\begin{eqnarray}
&&\sigma_{FSI}^j= {1\over 2} \sum\limits_n \delta(k_1+k_2-p_n)
[ M^*(n\to \pi^+\pi^-) T_j (B\to n)\nonumber \\
&&+ M (n\to \pi^+\pi^-)  T_j^* (B\to n) ],\quad j=u,c\,,
\label{GT}
\end{eqnarray}
where  $M(n\to \pi^+\pi^-)$ denotes the amplitude of the strong 
transition from the intermediate state $n$ to the final $\pi^+\pi^-$ state, 
and $T_j(B\to n)$ is the specific part of the weak decay amplitude of $B$ 
into the same intermediate state 
(divided by the constant factorized in Eqs. (\ref{abbns}) and (\ref{sigmaT})).

It is known that the strong dynamics at high energies is dominated by
multiparticle  production. However, as argued in  \cite{Zenc}, the
contribution of the multiparticle intermediate  states in  $B$ decay
is suppressed by  a flavour mismatch  between  the weak  and the strong
parts of the process. Therefore, only the  states composed
of two low mass resonances are expected to bring an important contribution 
to the rescattering. As shown in \cite{Zenc}, this picture is consistent
with the absence of final state interactions in $B$ decays in the heavy mass 
limit \cite{BeBu}, since the  production of two resonances is expected to 
vanish at high energies. 

In the two-particle approximation of the unitarity sum, the weak decay 
amplitudes
are completely specified by the masses of the particles in the intermediate
states, being independent on the phase space variables \cite{CaMi}. Then the
phase space integration implicit in Eq. (\ref{GT}) can be  performed exactly, 
leading to
\begin{eqnarray}  
&&\sigma_{FSI}^j= {1\over 2} \sum\limits_{\bar P_aP_a, \lambda} 
[M_{0, \lambda}^*
(\bar P_a P_a\to \pi^+\pi^-) T_j^\lambda (B\to \bar P_a P_a) \nonumber
\\ &&+M_{0, \lambda}  (\bar P_a P_a\to \pi^+\pi^-) T_j^{ \lambda *} 
(B\to \bar P_a P_a) ]\,+\ldots\,,
\label{GT2p}
\end{eqnarray}
where $M_{0,\lambda} (\bar P_a P_a\to \pi^+\pi^-)$ denote the $S$-wave 
projection of the  strong amplitudes and a summation over the 
helicities $\lambda$  of the intermediate states must be peformed in general.
By inserting this expression in Eq. (\ref{direlT}), and writing explicitly the 
real and the imaginary part of the logarithm, we obtain the relation 
\begin{eqnarray}
 &&T_j (B\to \pi^+\pi^-) = T_{j,0}(B\to \pi^+\pi^-) \nonumber\\
 &&+\left[\sum\limits_{\bar P_aP_a, \lambda} {\mbox Re}[ M_{0, \lambda}^*
(\bar P_aP_a\to \pi^+\pi^-) T_j^\lambda (B\to \bar
P_aP_a)]+\ldots\right] \nonumber\\ &&\times\left[i+{1\over \pi} \ln \left(1-
{m_\pi^2\over (m_B-m_\pi)^2}\right)\right]\,, \quad j=u,c\,, \label{direlT1}
\end{eqnarray}  
The sum includes the dominant two-particle quasielastic states and resonances 
$\bar P_aP_a$, and the  dots represent the contribution of the
multiparticle states. 

The strong amplitudes entering Eq. (\ref{direlT1}) are  evaluated at the  
c.m. energy squared equal to $m_B^2\approx 25 \,{\rm GeV}^2$. 
For low masses of the intermediate  particles 
we can use the generic Regge amplitude \cite{Collins}
\begin{equation}\label{regge0}   
-\gamma (t)\,{\tau + {\rm e}^{-i\pi\alpha(t)}\over \sin \pi\alpha(t)}\,
\left({s\over s_o}\right)^{\alpha(t)}\,, 
\end{equation} 
where $\gamma (t)$ is the residue function, $\tau$ the signature, 
$\alpha(t) =\alpha_0 +\alpha' t$ the linear trajectory, and  
$s_0\approx 1\, {\rm GeV}^2$. 
Modifications of the  standard Regge expression when the particles have 
larger   masses  are discussed in \cite{Zenc}. The $S$-wave projection of 
the amplitude (\ref{regge0}) in the spinless case is
\begin{eqnarray}\label{M_0}
M_{0, 0} (\bar P_a P_a\to \pi^+\pi^-)
 \approx \xi\, { \gamma(0)  \over 32
\pi\alpha'}\,{1\over  m_B
q L }\nonumber\\\times {\rm e}^{(\alpha_{0} +\alpha' t_0+2\alpha' q q') L}
\left[1- {\rm e}^{-2\alpha' q q' L}\right]\,, 
\end{eqnarray} 
where 
$q=1/2\sqrt{m_B^2-4m_\pi^2}$ and $q'=1/2\sqrt{m_B^2-4m_a^2}$ are the c.m.s. 
momenta of the final and intermediate state, respectively, $t_0=-(q^2+q'^2)$, 
$L=\ln m_B^2/s_0-i\pi/2$ and $\xi$ is a signature factor (equal, in particular,
to $-1$ for the pomeron and $i\sqrt{2}$ for the $\rho$ trajectory 
\cite{CaMi}).  In deriving Eq. (\ref{M_0}) we neglected the $t$ dependence 
of the ratio $\gamma(t)/\sin(\pi\alpha(t)/2)$ for $\tau=1$ trajectories, 
and of the ratio $\gamma(t)/\cos(\pi\alpha(t)/2)$ for $\tau=-1$.

\section{ Quark-hadron duality}\label{sec4}
The relation between the QCD predictions and the
hadronic physics is an extremely complex, still unsolved problem.
For testing quark-hadron duality it is in principle necessary to perform an
analytic continuation from the spacelike region of momenta, where
Operator Product Expansion and perturbative QCD are valid, to the timelike
axis, where the physical processes are  described in terms of hadronic
degrees of freedom. This procedure, based on dispersion relations, was
applied to simple objects, like the current-current vacuum correlation
functions or the electromagnetic form factors.

The weak hadronic decays are much more complicated, due to the presence of
hadrons in both initial and final states. Usually, the strong processes at 
scales larger than $m_b$ are integrated out, being included in the Wilson 
coefficients entering the effective hamiltonian (\ref{Hw}).  In perturbative
QCD, the decay amplitudes are treated in the heavy mass limit using the
framework of perturbative factorization for exclusive processes, based on
hard scattering kernels and light-cone  distribution amplitudes,  with the 
heavy mass playing the same role as the large momentum transfer.

In the hadronic picture, we shall consider the dispersion relation
(\ref{direlT}) (written in more detail in (\ref{direlT1}))  where, as
discussed in Section 2,  the first term $T_{j,0}$  is given by the
contributions which remain after switching off the final state strong
interactions among the emitted pions. This term should be provided, in
principle, by a nonperturbative calculation, which excludes in a systematic
way the final state interactions. As such a calculation is lacking, we
resort to a qualitative discussion, based on quark-hadron duality. 

We recall that, according to the discussion below Eq. (\ref{direla}), 
the terms $T_{j,0}$ include the degenerate terms and the
last two dispersive integrals  in the relation (\ref{inta2}). In
all these terms, the two final pions appear in different matrix elements, and
may be associated qualitatively to diagrams with no  gluon exchanges between
them. In particular, in the spirit of the dispersive formalism, the last two
integrals in Eq. (\ref{inta2}) may be interpreted as "initial state
interactions", in a crossed channel. Therefore, it is reasonable to assume 
that a considerable part of $T_{j,0}$ consists of the
naive factorized amplitude, which is  associated to processes with no gluon
exchanges between the emitted pion $\pi^+$ and the system $(\pi^-B_d^0)$ 
(except those already included  in the Wilson coefficients and the short 
distance processes taking place before hadronization). As for the
last term in the relation (\ref{direla}) (or (\ref{direlT1})), which  
describes in the hadronic picture the final state strong interactions, it is
dual to  the topologies (penguin annihilation, exchange diagrams,
scattering of the spectator quark, vertex corrections to the emission
diagrams), involving gluon exchanges between the final pions. Of course, the
correspondence between various quark diagrams and the terms appearing in the
hadronic formalism is not simple beyond the lowest orders of perturbation
theory. 
In particular, it is impossible to associate in an univoque way the
diagrams  involving many gluons either to the initial, or the final state
interactions. Moreover, from the point of view of final state interactions,
the spectator quark does not play a special role.  Therefore, the vertex
corrections involving the quarks emitted in the weak process, which  are
included in the  factorized part of the amplitude in the standard QCD 
factorization approach \cite{BeBu}, contribute also  to the final state 
interaction part of the amplitude. 

In the QCD factorization approach \cite{BeBu}, the  dominant 
contribution to the decay amplitude is given by the factorized term,  
with corrections which are suppressed, in the heavy limit $m_b\to \infty$,  
either by powers of $\alpha_s(m_b)$, or by powers of $\Lambda_{QCD}/m_b$. 
As discussed recently, the power suppressed corrections might be
enhanced by pure soft effects, such as endpoint singularities and
higher twist terms in the pion distribution amplitudes, appearing in   
annihilation diagrams or in the hard spectator interactions.
A recent evaluation of these corrections in perturbative QCD factorization
was done in \cite{BeBu2}, but there are still differences between the 
results obtained by different authors. 

In what follows we shall estimate the heavy mass corrections produced by 
the final state interactions, using the hadronic dispersion relation written 
in the form (\ref{direlT1}).
We shall consider first the contribution of the elastic and quasielastic
rescattering, taking as intermediate states $\bar P_aP_a$
the lowest pseudoscalar mesons  $\pi^+\pi^-$,  $\pi^0\pi^0$, $\bar K^0 K^0$, 
$K^+ K^-$ and $\eta\eta$. The effect of higher resonances describing
inelastic rescattering will be discussed below.

We assume that the amplitudes $T_j$ appearing in Eq. (\ref{direlT1}) can
be expanded in the heavy quark limit, {\it i.e.} for large $m_B\approx m_b$,as 
\begin{equation}\label{hlT} T_j\approx T_{j,0} + O(\alpha_s(m_B))
+O(\Lambda/m_B)+ \ldots\,, 
\end{equation} 
where $T_{j,0}$ are approximately given by the factorized amplitudes, with 
short distance corrections. These values have small imaginary parts, 
produced by the complex effective Wilson coefficients, vertex corrections or
short distance effects in the penguin and annihilation diagrams.

It is easy to write down the high energy limit of all the quantities entering
Eq. (\ref{direlT1}). 
First, from the explicit expression of the logarithm in the r.h.s. of this
relation, it follows  that the real part of FSI amplitude is suppressed by two
powers of the heavy mass with respect to the imaginary part. The heavy limit
behaviour of the Regge amplitude (\ref{M_0}) depends of the specific
trajectory. For the pomeron  (with $\alpha_0\approx 1.0$ and $\alpha'
\approx 0.25$) we obtain the expansion 
\begin{equation}\label{hlpom}
M_{0}^{(P)} \approx\, {\gamma_P(0)  \over 4 \pi }\left[{i\over \ln {m_B^2\over
s_0}}- {\pi\over 2} {1\over \ln^2{m_B^2\over s_0}}\,+\ldots\,\right]\,,
\end{equation} 
which shows that the dominant contribution in the heavy mass limit is 
imaginary. At the physical scale, using $\gamma_P(0) \approx 25.6$ 
\cite{CaMi}, we obtain $M_{0}^{(P)}\approx -0.23+0.69 i$.

For a physical trajectory, like $\rho$, using $\alpha_0\approx 0.5$ 
and $\alpha' \approx 1$, we obtain
\begin{equation}\label{hlphys}
 M_{0}^{( \rho)} \approx { \gamma_\rho(0)  \over 16 \pi m_B }
\left[{i+1\over \ln {m_B^2\over s_0}}-
{\pi\over 2} {1-i\over \ln^2{m_B^2\over s_0}}\,+\ldots\,\right]\,.
\end{equation}  
This amplitude is suppressed by one power of $m_B$ compared to the pomeron
amplitude (\ref{hlpom}). At the physical scale, using
$\gamma_\rho(0) \approx 31.4$ \cite{CaMi},
we obtain for the $\rho$ trajectory $M_{0}^{\rho} \approx 0.015+0.047 i$.
The masses of the particles undergoing the strong scattering appear 
as power suppressed terms in the expansions (\ref{hlpom}) and 
(\ref{hlphys}).

By inserting the expansions (\ref{hlT}) - (\ref{hlphys}) in
 Eq. (\ref{direlT1}), we can derive iteratively the magnitude of the 
coefficients of the  logarithmic and power corrections in the heavy limit 
expansion (\ref{hlT}).  As an illustration of the method, we take as input 
of the iterative procedure the values
\begin{eqnarray}\label{T0} 
T_0= T_{u,0}&=& 0.969-0.017 i\nonumber\\
P_0= {T_{c,0}\over R_b}&=&0.246+0.03 i\,,
\end{eqnarray}
which are typical for the factorized amplitude with short distance 
corrections   \cite{BeBu2}.

Using these values as the lowest approximation of the amplitudes $T_j$ in the 
right hand side of the relation (\ref{direlT1}), we obtain to first order
\begin{eqnarray}\label{Tout}
T=T_{u,1}&\approx & 0.969- 0.23 i \nonumber\\
P= {T_{c,1}\over R_b}&\approx &0.246 + 0.00012 i \,.
\end{eqnarray} 
The dominant corrections are given by the pomeron contribution to the 
elastic channel.
In particular, the imaginary part of $T_{u,1}$ is due mainly to the next to
leading order (second)  term in the expansion (\ref{hlpom}) of the
pomeron amplitude, since the dominant term is suppressed
by the symmetric Goldberger-Treiman summation in Eq. (\ref{direlT1}). 

The values (\ref{Tout}) have uncertainties due to the higher order terms in
the  heavy mass expansion. First we notice that the contribution of the
pseudoscalar mesons $\pi^0\pi^0$, $\bar K^0 K^0$, $K^+ K^-$ and $\eta\eta$,
responsible for the quasielastic rescattering in the unitarity sum of 
Eq. (\ref{direlT1}), is negligible. Indeed,  these states are produced 
by non dominant decay diagrams, and the corresponding  Regge amplitudes 
are described by physical trajectories, which are suppressed with respect 
to the pomeron.  The estimates made in  Ref. \cite{CaMi}, based on $SU(3)$ 
flavour symmetry,  show that the effect of these channels on the spectral 
functions is not larger than several percents.

The inelastic states might bring an important contribution in the unitarity
sum if they are produced by dominant decay diagrams, or the 
corresponding CKM coefficients are large. 
The first states which can contribute are the lowest  vector resonances, 
$\rho^+\rho^-$, $ \rho^0\rho^0$, $\bar K^* K^*$, $ \omega \omega$ and 
$ \phi \phi$, which describe  intermediate states with four or six pions.
The quark diagrams for $B$ decays into $VV$ pairs are 
similar to those  of the corresponding pseudoscalar mesons.  
Therefore, only the pair $\rho^+\rho^-$  is produced by a dominant tree
diagram. Estimates based on factorization \cite{AlKr} give for the helicity
amplitudes of the weak decay $B^0_d\to \rho^+\rho^-$ values comparable, up 
to a factor of 2, with the amplitude of the decay  $B^0_d\to \pi^+\pi^-$.  
As concerns the strong amplitude $M(\rho^+\rho^- \to  \pi^+\pi^-)$, 
it is described in the Regge model by the
exchange of $\omega$  and $A_2$, with  trajectories $\alpha\approx 0.5 +  t$ 
(almost degenerate with that of $\rho$), and the $\pi$ exchange,
whose trajectory $\alpha_\pi \approx 0. + 0.9t$ is non-dominant
\cite{Collins}.  
As discussed in Ref. \cite{Jones}, the kinematic factors of the $t$-channel 
helicity amplitudes suppress in the present case the contribution of the  
natural parity exchanges $\omega$  and $A_2$, so that the scattering amplitude 
of $\rho^+\rho^- \to  \pi^+\pi^-$ is described at small $t$ by 
$\pi$ exchange. This implies a considerable suppression, 
verified indirectly in the related reactions of vector meson
production $\pi^\pm N\to \rho^\pm N$ \cite{Pratt,FoQu}.
Therefore, the vector meson resonances seem to  bring a negligible
contribution to the inelastic rescattering in $B^0_d\to \pi^+\pi^-$ decay.
Using this spin suppression argument, we expect that the same conclusion 
applies also to  other vector mesons of higher mass.

The role of the intermediate states with charm, like the pair $\bar D D$, 
has been considered by several authors and is still controversial.
Since a large part of the inelastic $\pi\pi$ scattering at 
$\sqrt{s}\approx m_B$ goes into multiparticle states 
composed of noncharmed mesons, the contribution of these states
was assumed in  Ref. \cite{Zenc} to be negligible. 
The annihilation of the  $\bar c c$ pair in the  penguin diagrams
proceeds  therefore through a short distance interaction, 
which can be computed perturbatively.
We mention however that the possibility of a long distance 
contribution of the ``charming penguins'' was also 
considered by some authors  \cite{CiFr}, possibly through the intermediate 
state $\bar D D$ \cite{Kamal,Xing}. 
A detailed estimate is difficult since the validity of the Regge model is 
questionable at $\sqrt{s}\approx m_B$  for large masses 
($m_D=1.68\, {\rm GeV}$) of the particles undergoing the strong rescattering
\cite{Zenc}. We mention  that the contribution of high mass resonances, 
such as $\bar DD$, is actually suppressed by the phase space integral  
in Eq. (\ref{GT}), evaluated at fixed $s=m_B^2$. In the present 
approach, we assume also  that their global effect is taken into
account in a certain measure by the Goldberger-Treiman symmetric summation.
With this assumption, we do not expect other important power corrections in the
spectral functions. 

From the values given in Eq. (\ref{Tout}) we obtain
\begin{equation}\label{rap}
\left\vert{P \over  T }\right\vert \approx 0.246
\,,\quad  \delta={\rm Arg} \left[{P \over T}\right] \approx
14^\circ \,,  
\end{equation} 
while the input values (\ref{T0}) correspond to  $|P_0/T_0| = 0.25$
and ${\rm Arg} [P_0/T_0]= 8 ^\circ$. Thus, in the present approach the
final state interactions do not modify considerably the modulus and the
phase of the ratio  $P/T$. We recall that the lowest order values 
$T_{j,0}$ depend on the renormalization scale $\mu$ in the Wilson coefficients.
It is expected that this dependence will be diminished by the inclusion of 
other terms in the l.h.s. of  Eq. (\ref{direlT1}). 

\section{Conclusions}\label{sec5}
In the present work we investigated  the effects of the final state
interactions to $B_d^0\to \pi^+\pi^-$ decay in a formalism  based on hadronic
unitarity and dispersion relations. We discussed the heuristic derivation of
the dispersion relations with respect to the momenta of the external
particles, by applying the LSZ procedure to the $S$-matrix element of the
weak decay.  The ambiguities which affect in general the off-shell
extrapolation of the amplitudes, appear in our formalism in the so-called 
"degenerate terms", produced by the equal-time commutators in the LSZ 
formalism, and in the off-shell quantities entering the spectral functions. 

The evaluation of the dispersion relations in external momenta  is in 
general very complicated. However, we noticed that the dispersion relation 
with respect to the momentum squared of one final pion can be written in the  
convenient form (\ref{direla}), where the contribution of the final state 
interactions is separated from other terms, and involves  only on-shell 
quantities. We used this relation, written in more detail in 
Eq. (\ref{direlT1}), as an iterative scheme for determining the  
corrections to the factorized amplitude, generated by the final 
state interactions in the heavy mass limit.   
A nontrivial prediction of the formalism is that the real
part of  the FSI contribution is suppressed by two powers of the heavy mass,
compared to the imaginary part. Using for illustration a numerical input
suggested by QCD factorization, we noticed the dominant effect of the next to
leading  logarithmic term of the pomeron contribution. Other sources of large
power corrections to the factorized amplitude are not found.
We discussed the contribution of the lowest pseudoscalar mesons  and vector
meson resonances, and  assumed that the effects of higher resonances 
and multiparticle states are qualitatively taken into account by the
Goldberger-Treiman method of calculating the spectral functions. In
particular, the results of a numerical test indicate that the phase and the
modulus of the ratio $P/T$ are not drastically modified by the final state
interactions.  Using improved results of QCD calculations, 
it will be possible to test the dispersion relations conjectured 
in the present work and, more generally, the validity of quark-hadron duality.

\vskip 0.3cm  {\bf Acknowledgments:}  This work  was realized in the frame
of the Cooperation Agreements between IN2P3 and NIPNE-Bucharest, and  between
the CNRS and the Romanian Academy. Two of the authors (I.C. and L.M.) received
financial support from the Romanian National  Agency for Science, Technology
and Innovation, under the Grant Nr. 6126/2000.



\begin{thebibliography}{99}

\bibitem{Ball} P.Ball {\it et al.} ``B decays at the LHC'' hep-ph/0003238,
published in CERN Yellow Report CERN-2000-004, ``1999 CERN Workshop on Standard
Model Physics (and more) at the LHC'', G. Altarelli \& M. Mangano, eds.

\bibitem{CLEO} D. Cronin-Hennessy {\it et al.}, (CLEO Collaboration), 
Phys. Rev. Lett. {\bf 85} (2000) 515.

\bibitem{BaBar} G. Cavoto (BaBar Collaboration), 
 talk at the 36th Rencontres de Moriond on QCD and Hadronic
Interactions, Les Arcs, France, 17-24 March 2001, to appear in the proceedings
, hep-ex/0105018. 

\bibitem{Belle} T. Iijima (Belle Collaboration), 
talk at the
 4th International Conference on B Physics and CP
Violation (BCP 4), Ago Town, Japan, 19-23 February 2001, to 
appear in the proceedings, hep-ex/0105005. 


\bibitem{BaSt} M. Bauer, B. Stech and M. Wirbel, Z. Phys. C {\bf 34} 
(1987) 103; Z. Phys. C {\bf 29} (1985) 637.

\bibitem{DuGr} M.J. Dugan and B. Grinstein, Phys. Lett. B {\bf 255} (1991) 583.

\bibitem{DoPe} J.F. Donoghue and A.A. Petrov, Phys. Lett. B {\bf 393}
(1997) 149.

\bibitem{AlGr} A. Ali and C. Greub, Phys. Rev. D {\bf 57} (1998) 2996.

\bibitem{BeBu} M. Beneke, G. Buchalla, M. Neubert and C.T. Sachrajda,
Phys. Rev. Lett. {\bf 83} (1999) 1914;   Nucl.
Phys. B. {\bf 591} (2000) 313.

\bibitem{BeBu1} M. Beneke, G. Buchalla, M. Neubert and C.T. Sachrajda,
Contribution to ICHEP2000, July 27 - August 2, Osaka, Japan, PITHA-00-13,
[hep-ph/0007256].

\bibitem{Bene} M. Beneke, PITHA 00-23, [hep-ph/0009328].

\bibitem{DuYa} D. Du, D. Yang and G. Zhu, Phys. Lett. B. {\bf 488} (2000) 46.

\bibitem{MuSu} T. Muta, A. Sugamoto, M.Z Yang and Y.D. Yang, Phys. Rev. D 
{\bf  62} (2000) 094020.

\bibitem{BeBu2} M. Beneke, G. Buchalla, M. Neubert and C.T. Sachrajda,
 hep-ph/0104110.

\bibitem{KeLi} Y.Y. Keum, H. Li and A. Sanda, Phys. Rev. D {\bf 63} (2001) 
054008; Y.Y. Keum and H. Li, Phys. Rev. D {\bf 63} (2001) 074006.

\bibitem{LuUk} C.D. Lu, K. Ukai and M.Z. Yang, Phys. Rev. D {\bf 63} 
(2001) 074009.

\bibitem{Dono} J.F. Donoghue {\it et al.}, Phys. Rev. Lett. {\bf 77}
(1996) 2178.

\bibitem{BlHa} B. Blok and I. Halperin, Phys. Lett. B {\bf 385} (1996) 324.

\bibitem{Dono1} J.F. Donoghue, E. Golowich, and A.A. Petrov, Phys. Rev. D
{\bf 55} (1997) 2657.

\bibitem{BlGr} B. Blok, M. Gronau and J.L. Rosner, Phys. Rev. Lett.
{\bf 77} (1997) 3999.

\bibitem{Falk} A.F. Falk, A.L. Kagan, Y. Nir, and A.A. Petrov, Phys. Rev. D
{\bf 57} (1998) 4290.

\bibitem{Dele} D. Del\'epine, J.-M. G\'erard, J. Pestieau, and J. Weyers, Phys.
Lett. B {\bf 429}  (1998) 106; J.-M. G\'erard, J. Pestieau, and J. Weyers,
Phys. Lett. B {\bf 436} (1998) 363.

\bibitem{Kamal} A. N. Kamal,  Phys. Rev. D {\bf 60}
 (1999) 094018.

\bibitem{Xing} Z. Xing, Phys. Lett. B {\bf 493} (2000) 301.

\bibitem{CaMi} I. Caprini, L. Micu, and C. Bourrely, Phys. Rev. D {\bf 60}
 (1999) 074016; D {\bf 62} (2000) 034016.

\bibitem{SuWo} M. Suzuki, L. Wolfenstein, Phys. Rev. D {\bf 60}
(1999) 074019.

\bibitem{Zenc} P. Zenczykowski,  Phys. Rev. D {\bf 63}
(2001) 014016; hep-ph/0102186.

\bibitem{LSZ} H. Lehmann, K. Symanzik, and W. Zimmermann, Nuovo Cimento,
{\bf 1}, (1956) 205; {\bf 2}, (1957) 425.

\bibitem{GoWa} M. L. Goldberger and K. M. Watson, \textit{Collision Theory},
(Wiley, New York, 1964).

\bibitem{Bart} G. Barton, 
\textit{Introduction to Dispersion Techniques in Field Theory},
 (Benjamin, New York, 1965).

\bibitem{Binc} A.D. Bincer, Phys. Rev. {\bf 118} (1960) 855.

\bibitem{Suzu} M. Suzuki, hep-ph/0102028.

\bibitem{KaWi} G. K\"alen and A. S. Wightman, K. Dan. Vidensk. Fys. Skr. 
{\bf 1} (1958) No. 6.

\bibitem{PaPi} E. Pallante and A. Pich, Phys. Rev. Lett. {\bf 84} (2000) 2568.

\bibitem{Buras} A. J. Buras et al, Nucl. Phys. B{\bf 565} (2000) 3.

\bibitem{BuCo} M. B\"ucher, G. Colangelo, J. Kambor and F. Orellana,
hep-ph/0102387, hep-ph/0102389.

\bibitem{GoTr} M.L. Goldberger and S.B. Treiman, Phys. Rev.
{\bf 110} (1958) 1178; {\bf 111} (1958) 354.

\bibitem{MaTe} L. Maiani and M. Testa, Phys. Lett. B {\bf 245} (1990) 585. 

\bibitem{Collins} P.D.B. Collins,
\textit{Introduction to Regge theory and High Energy Physics}, 
(Cambridge University Press, Cambridge, England, 1977).

\bibitem{AlKr} A. Ali, G. Kramer and C. D. L\"u, 
Phys. Rev. D {\bf 58} (1998) 094009.

\bibitem{Jones} L. Jones, Phys. Rev. {\bf 163} (1967) 1523.

\bibitem{Pratt} J. C. Pratt {\it et al.}, Phys. Lett. B {\bf 41} (1972) 383.

\bibitem{FoQu} G. F. Fox and C. Quigg, Ann. Rev. Nucl. Sci (1973) 219.

\bibitem{CiFr} M. Ciuchini, E. Franco, G. Martinelli and L. Silvestrini, 
Nucl. Phys. B {\bf 501} (1997) 271;  M. Ciuchini {\it et al.}, hep- ph/0104126.

\end{thebibliography}
\end{document}